\documentclass[a4paper]{article}%
\usepackage{algorithm,algpseudocode}
\usepackage{graphicx}
\usepackage{fullpage}
\usepackage{amsfonts,amsmath,amssymb,amsthm, bm}

\begin{document}

\title{Challenges and Insights: Exploring 3D Spatial Features and Complex Networks on the MISP Dataset}
\author{Yiwen Shao\\\\
  Center for Language and Speech Processing,\\Johns Hopkins University, 
  \\Baltimore, MD, USA \\
  \small \tt yshao18@jhu.edu}
\date{Sep, 2022 - May, 2023}
\maketitle

\section{Abstract}
Multi-channel multi-talker speech recognition presents formidable challenges in the realm of speech processing, marked by issues such as background noise, reverberation, and overlapping speech. Overcoming these complexities requires leveraging contextual cues to separate target speech from a cacophonous mix, enabling accurate recognition. Among these cues, the 3D spatial feature has emerged as a cutting-edge solution, particularly when equipped with spatial information about the target speaker. Its exceptional ability to discern the target speaker within mixed audio, often rendering intermediate processing redundant, paves the way for the direct training of "All-in-one" ASR models. These models have demonstrated commendable performance on both simulated and real-world data. In this paper, we extend this approach to the MISP dataset to further validate its efficacy. We delve into the challenges encountered and insights gained when applying 3D spatial features to MISP, while also exploring preliminary experiments involving the replacement of these features with more complex input and models.
\section{Introduction}
\label{sec:intro}
Multi-channel multi-talker speech recognition continues to pose significant challenges within the speech processing community. In contrast to clean, non-overlapping single-speaker speech, multi-talker speech frequently contends with issues such as background noise, pronounced reverberation, and overlapping interfering speech. Addressing this complexity necessitates the provision of additional contextual information to facilitate the separation of the target speech from the noisy mixture, subsequently enabling accurate speech recognition. Various cues have been explored for this purpose, including speaker identification, lip embeddings, face embeddings, and spatial features, among others.

Among these cues, the 3D spatial feature stands out as a state-of-the-art feature, particularly when working with multi-channel recordings and possessing spatial information about the target speaker. Thanks to its robust ability to discriminate the target speaker within the mixed time-frequency (T-F) bins, even intermediate speech separation or enhancement modules become unnecessary. As a result, an "All-in-one" ASR model can be trained directly, achieving comparable performance on both simulated data and real-world in-car data, as demonstrated in \cite{shao2022multi}. In light of these promising outcomes, this work applies the same approach to the MISP dataset to further validate its suitability and effectiveness.

In the following sections we will describe the challenges and insights we have when we apply 3D spatial feature on the MISP dataset. Soe prmelimary experiments of replacing 3D spatial feature with complex input and models are also discussed.

\section{Review on 3D Spatial Feature}

The Spatial Feature (SF) or angle feature (AF), as previously introduced in \cite{gu2019neural, shao2022multi}, serves the purpose of indicating the dominance of the target source within the time-frequency (T-F) bins. We define a target-dependent phase difference (TPD) as the phase delay induced by a theoretical wave, with frequency $f$, traveling from the target's location $\mathbb L_t$ to be measured at the $m$-th microphone pair at time $t$, as described in \cite{gu2020multi}. In general, the spatial features pertaining to a target speaker are computed as the cosine distance between the \textbf{target-dependent phase difference (TPD)} and the \textbf{interchannel phase differences (IPD)}:

\begin{equation}
    \mathbb{SF}_{t,f} =\sum_{p=1}^P\langle {\bm e}^{\text{TPD}_{t,f}^{(m)}(\mathbb{L})}, {\bm e}^{\text{IPD}_{t,f}^{(m)}}\rangle
\end{equation}
where ${\bm e}^{(\cdot)}=\begin{bmatrix} \cos(\cdot) \\ \sin(\cdot)\end{bmatrix}$ and $\text{TPD}_{t,f}^{(m)} (\mathbb L)$ depends on target speaker's location $\mathbb L$ obtained from the vision. 

It calculates the phase difference at microphone pair $(m)$ corresponding to the presence of a wave originating from location $\mathbb L$. In simple terms, if a speaker at location $\mathbb L$ is actively speaking, the calculated "theoretical" phase difference $\text{TPD}_{t,f}^{(m)} (\mathbb L)$, as derived from visual information, will closely resemble the "observed" phase difference $\text{IPD}_{t,f}^{(m)}$. Consequently, the resulting $\mathbb{SF}_{t,f}$ value will tend to approach 1. Conversely, if there is no such speaker activity, the similarity between these phase differences diminishes, leading to a $\mathbb{SF}_{t,f}$ value closer to 0.

\vspace{-0.1cm}
\subsection{Target Phase Difference in 3D Space}
\begin{figure}[htb]
    \centering
    \includegraphics[width=9cm]{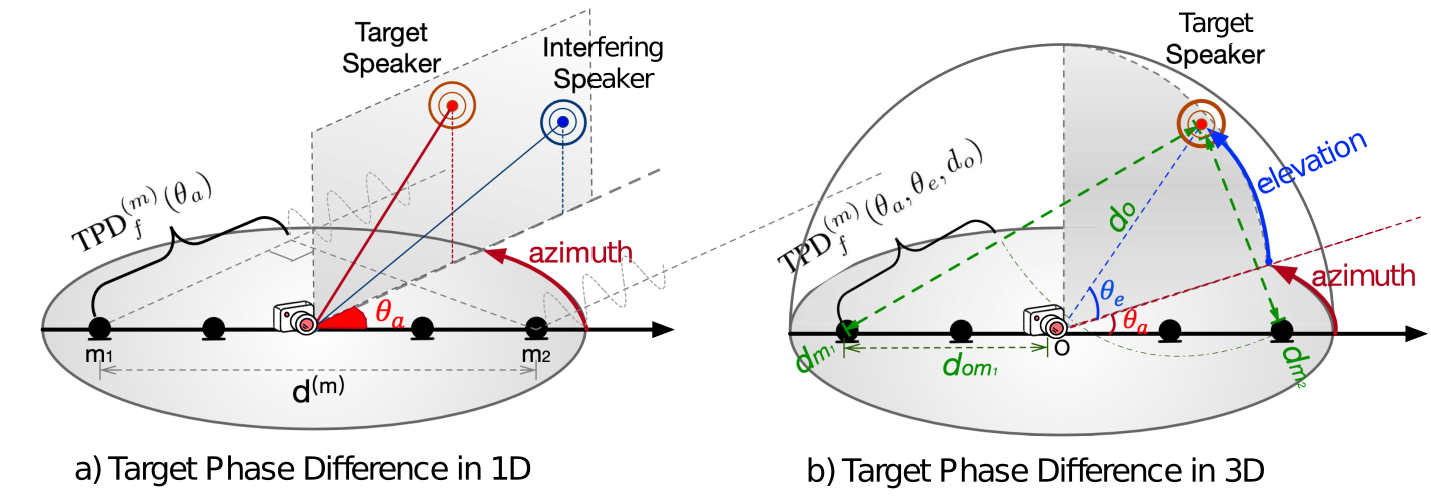}
    \vspace{-0.5cm}
    \caption{The illustration of 1D and 3D scenarios. Fig~\ref{fig:3D} a) shows inseparable issue when azimuths of two simultaneous speech are close.} \vspace{-0.3cm}
    \label{fig:3D}
\end{figure}

As our empirical study \cite{chen2018multi, gu2019neural} has revealed, when speakers are in close proximity, for example, with an azimuth difference (AD) of less than 15 degrees, the spatial information calculated solely based on azimuth is insufficient for effective speaker distinction. Consequently, there arises a necessity to extend the target-dependent phase difference (TPD) into 3D space, incorporating comprehensive location information denoted as $\mathbb L=[\theta_a, \theta_e, d_o]$. This expansion is visually depicted in Fig.~\ref{fig:3D}, and the resulting generalized $\text{TPD}_{t,f}^{(m)} (\mathbb L)$ in three dimensions is represented as:
\begin{equation}
\begin{split}
   & \text{TPD}_{t,f}^{(m)} (\theta_a, \theta_e, d_o) = \frac{2\pi f}{c(F-1)} \cdot f_s \cdot (d_{m_1} - d_{m_2}) \\
   &     d_{m_i} = \sqrt{d^2_{om_i}+d^2_o-2d_{om_i}d \cos \theta_a \cos \theta_e} ~~_{\forall i \in \{1,2\}}
\end{split}
\label{eq:TPD3D} \vspace{-0.1cm}
\end{equation}
where $\theta_a$ is azimuth, $\theta_e$ is elevation. $d_o$, $d_{m_1}$ and $d_{m_2}$ are distances between the target speaker and the camera, the $m_1$-th and the $m_2$-th microphone, respectively. $d_{om_i}$ is the distance between the camera and the $m_i$-th microphone (as shown in Fig.~\ref{fig:3D}). Here the camera is at the center of microphone array.

\section{Empirical Study On the Real-World Dataset: MISP}

\subsection{Dataset}
The MISP2021-AVSR dataset \cite{chen2022audio} is a real recorded audio-visual dataset, which comprises a total of 122.53 hours of audio-visual content, divided into 376 sessions. Each session encompasses approximately 20 minutes of discussion. In total, there are 248 distinct speakers featured in the dataset. The recordings were conducted in 30 real living rooms, varying in size from 3.2 × 2.56 × 2.54 to 5.2 × 4.2 × 2.8 cubic meters.

\subsection{System Overview}
A pruned version of RNNT model \cite{kuang2022pruned} is used as the ASR backbone.  The network is formed by a 12-layer 4-head Conformer \cite{gulati2020conformer} encoder with 512 attention dimensions and 2048 feed-forward dimensions. Log Mel-Filterbank (LFB) acoustic features with
40 bins are used, which are extracted with a 25ms window, 10ms hop-length at a sample rate of 16kHz.

\begin{table}[h]
    \centering
\begin{tabular}{l|l|l|r|r}
\hline Training Data & Input & Front-end & Near & Far \\
\hline near & audio & - & 24.65 & 83.60 \\
far & audio & - & 72.74 & 73.25 \\
far & audio & WPE + BeamformIT & - &$59.96$ \\
near+middle+far & audio & - & 21.87 & $54.70$ \\
near+middle+far & audio & WPE + BeamformIT & 22.17 & $54.77$ \\
\hline
\end{tabular}
\caption{CER (\%) of RNNT model on Near/Far field scenarios trained with different splits.}
\end{table}

Table 1 illustrates that employing WPE and BeamformIT as pre-processing front-ends offers clear advantages when training and testing the model solely with far-field overlapped speech. However, when the model's training data includes clean near-field and middle-field speech in addition to far-field speech, the importance of the front-end diminishes. Consequently, for the sake of simplicity, we exclude this module from our subsequent experiments.

Another noteworthy observation from Table 1 is that training the model with clean, non-overlapped speech alongside noisy far-field speech yields notable benefits. This approach helps the model prioritize the ASR task, resulting in improved checkpoint convergence. This focus on ASR, rather than being distracted by underlying separation and enhancement tasks due to the absence of specific separation or enhancement modules, aligns with findings from other studies \cite{xu2022channel, wangxiaomi} on this dataset. These studies have demonstrated that extensive data augmentation, involving both simulated overlapped speech and clean enhanced speech, significantly enhances ASR performance on far-field data.

\subsection{Usage of 3D Spatial Feature}
Since the MISP dataset includes video data along with comprehensive specifications of the microphone array, camera setup, and room layout, we can extract spatial information (including azimuth, elevation, and the target speaker's distance from the center of microphones) using a facial detection algorithm applied to the video footage. This enables us to harness the 3D spatial feature for analysis on this dataset, and the outcomes are presented in Table 2.

\begin{table}[h]
    \centering
\begin{tabular}{l|l|l|r}
\hline Training Data & Input & Front-end & Far \\
\hline far & audio & - & 73.25 \\
far & audio + video & - & 64.37 \\
far & audio + video & WPE + BeamformIT &58.28 \\
\hline
\end{tabular}
\caption{CER (\%) of RNNT model on MISP Far-field speech.}
\end{table}

Table 2 reveals that the inclusion of the 3D spatial feature extracted from the video significantly aids the model in distinguishing the primary target speaker from interfering speakers, leading to a notable reduction in CER from 73.25\% to 64.37\%. However, when the WPE + BeamformIT preprocessing module is already in place, which essentially serves a similar purpose, the additional 3D spatial feature from the video side offers comparatively limited improvement (59.96\% vs. 58.28\%).

It's worth noting that the requirement for extra spatial information renders it impractical to employ near-field and middle-field data during training. Consequently, this limitation results in a less favorable overall performance on far-field data.

\subsection{Swap Training}
To address the challenge of utilizing pure audio data that lacks spatial information, we propose an innovative training strategy designed to accommodate both scenarios—with and without video cues. We introduce two distinct branches for the input side of the model: one equipped with additional features, such as the 3D spatial feature, and the other relying on fundamental acoustic features, such as fbank representations. These two branches project the input data into the same hidden dimension, enabling them to share subsequent model components and receive updates collectively.

During the training process, audio inputs are randomly sampled from the dataset, forming batches that consist either entirely of extra features or solely of standard acoustic features. A hyperparameter denoted as $\alpha$ is employed to regulate the proportion of batches from each branch, aiming to optimize model performance on the target test set. In our experiments, $\alpha = 0.7$ is found to be the best choice.

\begin{table}[h]
    \centering
\begin{tabular}{l|l|l|r}
\hline Training Data & Input & Front-end & Far \\
\hline far & audio & - & 73.25 \\
far & audio + video & - & 64.37 \\
far & audio + video & WPE + BeamformIT &58.28 \\
near + middle + far & audio & WPE + BeamformIT & 54.77 \\
near (0.7) + far (0.3) & audio + video & WPE + BeamformIT & \textbf{50.03} \\
\hline
\end{tabular}
\caption{CER (\%) of RNNT model on MISP Far-field speech.}
\end{table}

\subsection{Problem of 3D Spatial Feature: Reverberation}
Both Table 2 and 3 demonstrate that the most substantial enhancement observed when adding the 3D spatial feature to the pure audio baseline is a reduction from 54.77\% to 50.03\%. This improvement falls notably short of our expectations, especially considering our prior experiments on simulated data and real-world in-car data \cite{shao2022multi}, where the 3D spatial feature typically yielded performance approaching the upper bounds set by clean, non-overlapped speech (ranging from 20\% to 30\% CER on the MISP dataset). To understand this discrepancy, we closely examine the feature maps of IPD, TPD, and $\mathbb{SF}$ on different simulated data, as depicted in Figure 2.

\begin{figure}[h]
\vspace{-0.1cm}
\centering
    \includegraphics[width=80mm, height=60mm]{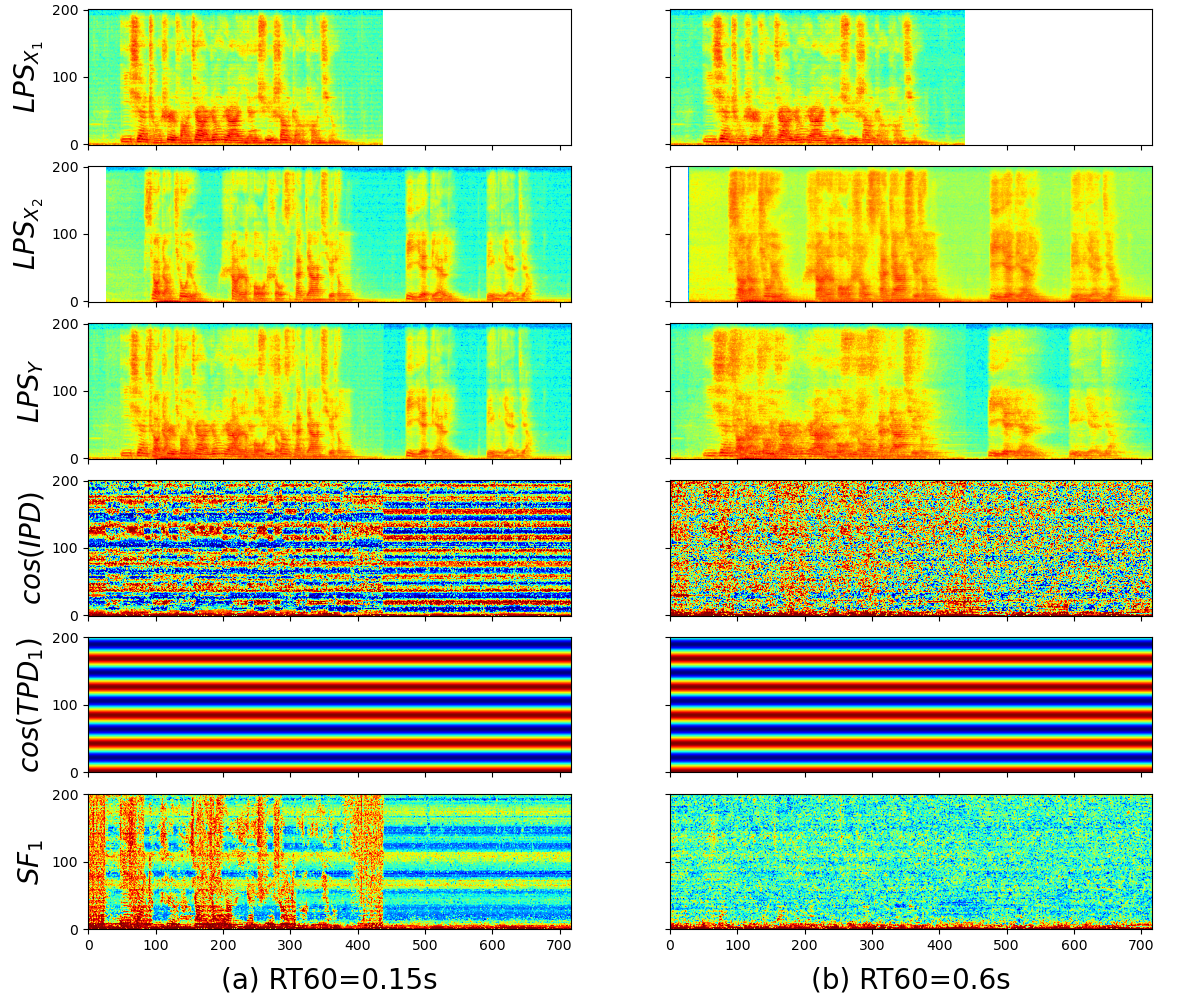}
    \caption{\small An illustration of 3D spatial features with (a) weak reverberation and (b) strong reverberation. In (a), $\mathbb{SF}_1$ matches well with the pattern of the Log Power Spectrum (LPS) of the target speaker $X_1$. While in (b), $\mathbb{SF}_1$ fails to identify the target source from the mixture.}\vspace{-0.4cm}
    \label{fig:3d_sf}
\end{figure}

Due to the presence of non-negligible reflection waves in the overlapped speech, IPD exhibit considerable noise and cannot serve as a reliable reference for comparison with the TPD. This results in an unfavorable feature map, as illustrated in Figure 2(b).

\subsection{3D Spatial Feature Replacement: Complex Input and Networks}
Due to the inherent influence of reverberation on the actual IPD captured by the microphones, as opposed to the theoretical TPD, the subsequent 3D spatial feature loses its ability to accurately indicate the dominance of the target speech within the time-frequency (T-F) bins.

Consequently, our approach is to directly employ the complex Short-Time Fourier Transform (STFT) signals as input, without segmenting them into magnitude and phase components. Instead, we construct a complex neural network designed to learn inherent patterns within the complex signals. To provide the model with additional spatial information regarding the target speaker, we introduce the target speaker's steering vector as a specific cue appended to the complex input. In concrete terms, the input and model architectures in the complex domain are as follows:

\textbf{Input:} The input consists of the concatenation of the covariance matrix of the multi-channel spectrogram and the steering vector, resulting in a tensor with dimensions $[B, 2MM, F, T]$, where B represents the batch size, M denotes the number of channels, F signifies the number of frequency bins, and T represents the number of frames.

\textbf{Model:} To ensure a fair comparison, our modifications are limited to the input embedding layer of the conformer encoder, while all other components remain unchanged. The initial input embedding layer comprises several \textit{conv2d} layers designed to transform input data, originally with a shape of $[B, I, T]$, into hidden embeddings with dimensions of $[B, H, T / 4]$, where $I$ represents the input dimension, and $H$ represents the hidden dimension. The sequence length is reduced by a factor of 4 due to the convolution layers' stride. However, since our multi-channel complex input is a 4-dimensional tensor, we implement the following modifications with 3 versions:

\noindent (1) \textit{Naive Approach}: In the naive version, the complex input is split into its Real and Imaginary parts, which are then concatenated along the channel dimension (i.e., $M^2$). This results in a new tensor with dimensions $[B, 4MM, F, T]$. Subsequently, this tensor is treated as if it were a real tensor in the subsequent layers, obviating the need for any special processing in this version.

\noindent (2) \textit{Separate Approach}: In the separate version, the complex input is divided into its Real and Imaginary parts, creating two distinct real tensors with dimensions $[B, 2MM, F, T]$ each. These real tensors are independently passed through the same convolutional blocks to produce two separate hidden representations, each with dimensions $[B, H, F, T]$. Finally, these two representations are fused by computing their magnitude: $Output = R^2 + I^2$.

\noindent (3) \textit{Cross-Product Approach}: Drawing inspiration from complex networks introduced in \cite{hu2020dccrn, gu2021complex}, we apply complex number multiplication rules during the forward pass of the network. The input is represented as [R, I], and the convolutional operation is performed as follows: Conv2d: $[W_r, W_i]$, Nonlinear: $F()$. The output is computed as $[F(W_r * R - W_i * I), F(W_i * R + W_r * I)]$. Similar to the separate version, the final output's real and imaginary parts are fused by calculating their magnitude.

We test these three approaches on the simulated Aishell dataset with the same settings in \cite{shao2022multi}. The results are shown in Table 4.

\begin{table}[h]
    \centering
\begin{tabular}{l|l|l}
\hline Approach & Dev & Test \\
\hline 3D SF & 10.76 & 12.48 \\
Naive &11.97 &14.61 \\
Separate & 19.05 & 23.92 \\
Cross-product* & 63.05 &90.3\\
\hline
\end{tabular}
\caption{CER (\%) of Complex RNNT models on simulated Aishell dev/test sets. * The bad results may come from poor convergence.}
\end{table}

Interestingly, none of the aforementioned complex approaches manage to surpass our prior best results achieved with the 3D spatial feature. Surprisingly, even the mathematically more precise version performs worse than its simpler counterparts. Our interpretation of these findings is that training the network directly on multi-channel complex input for the task of target speaker ASR proves to be a challenging endeavor, requiring additional human-crafted guidance for convergence.

In light of these results, we conclude that relying solely on a pure complex network with complex STFT input is insufficient for the network to effectively update its weights. Consequently, we believe that further efforts should be directed toward providing intermediate guidance to the model, rather than expecting it to learn entirely on its own.

\section{Conclusion and Future Work}
Through our examination of real-world challenge data from MISP, we have identified a critical limitation of the 3D spatial feature when confronted with significant reverberation—an aspect that had eluded our prior observations. Building upon this insight, we have taken a step back by relinquishing the use of spatial features entirely and replacing them with pure complex input and networks.

Our experimental results on simulated data suggest that wholly discarding all human-crafted features and expecting the model to autonomously learn representations may not be an advisable approach, particularly when data availability is limited. A more promising avenue may involve a reevaluation of spatial feature design or exploring alternative sources of spatial information, such as the \textbf{room impulse response}.

\bibliographystyle{IEEEbib}
\bibliography{main}

\end{document}